# Effect of Device Mismatches in Differential Oscillatory Neural Networks

Jafar Shamsi, María José Avedillo, Bernabé Linares-Barranco, *Fellow, IEEE, and* Teresa Serrano-Gotarredona

*Abstract*—**Analog implementation of Oscillatory Neural Networks (ONNs) has the potential to implement fast and ultra-low-power computing capabilities. One of the drawbacks of analog implementation is component mismatches which cause desynchronization and instability in ONNs. Emerging devices like memristors and $VO_2$ are particularly prone to variations. In this paper, we study the effect of component mismatches on the performance of differential ONNs (DONNs). Mismatches were considered in two main blocks: differential oscillatory neurons and synaptic circuits. To measure DONN tolerance to mismatches in each block, performance was evaluated with mismatches being present separately in each block. Memristor-bridge circuits with four memristors were used as the synaptic circuits. The differential oscillatory neurons were based on $VO_2$-devices. The simulation results showed that DONN performance was more vulnerable to mismatches in the components of the differential oscillatory neurons than to mismatches in the synaptic circuits. DONNs were found to tolerate up to 20% mismatches in the memristance of the synaptic circuits. However, mismatches in the differential oscillatory neurons resulted in non-uniformity of the natural frequencies, causing desynchronization and instability. Simulations showed that 0.5% relative standard deviation (RSD) in natural frequencies can reduce DONN performance dramatically. In addition, sensitivity analyses showed that the high threshold voltage of VO2-devices is the most sensitive parameter for frequency non-uniformity and desynchronization.**

## I. Introduction

Inspired by the synchronization phenomenon in the biological brain, oscillatory neural networks (ONNs) are processing architectures consisting of oscillators coupled with synaptic circuits [1]. In ONNs, data is encoded in the oscillator phase and processed through oscillator phase synchronization and phase difference. In ONNs with binary values, oscillators synchronize either in-phase or anti-phase, which are equivalent to values 1 and 0. Data encoding in the oscillator phase allows low amplitude oscillators to be used for processing, resulting in low power computations. Emerging low power and nano-scale devices like vanadium dioxide ($VO_2$) [2] and memristor [3] also make it feasible to implement oscillators and synaptic circuits efficiently. For instance, a basic oscillator circuit can be made from the series connection of a $VO_2$ device and a resistor or a CMOS transistor [4]. As a coupling component, a memristor can be used to interconnect the oscillators. Differential ONNs (DONNs) have recently been introduced that use differential oscillators, each providing one in-phase and one anti-phase signal [5]. Differential outputs allow a memristor-bridge synaptic circuit to be used to implement positive, negative, or zero weights [6].

Although the use of emerging devices is efficient as a means of implementing ONNs in hardware, variability in these components constitutes a serious constraint. $VO_2$ devices, for example, are prone to device-to-device and cycle-to-cycle variability in resistances and threshold voltages [4], [7]. This variability is the main problem for large-scale implementations of coupled oscillators [4], [8]. On the other hand, memristors also suffer from device-to-device and cycle-to-cycle variations [9], [10]. While device-to-device variability is related to fabrication processes such as differences in film thicknesses [10], cycle-to-cycle variations are associated with the memristor's random physical operating mechanisms, such as conductive filament variations [11]. Variability in basic components (resistors and capacitors) is inevitable, not only in emerging devices but also even in modern technologies. It is therefore essential to consider the device variability of all components in an ONN's hardware.

It is worth mentioning that theoretical analysis shows that ONNs require quasi-perfect synchronization for computing purposes [12], [13], which can be affected by device mismatches. For instance, in [14], it is shown that the synchronization of spin torque oscillators (STOs) is sensitive to device variations. In [15], [16], the robustness to noise and devices' non-idealities is studied in ONNs architectures that are based on the Kuramoto model (sinusoidal oscillators). To the best of our knowledge, a precise study of device mismatches in $VO_2$-based ONNs has not been provided in the literature.

This paper studies the effect of device variability (mismatches) on DONN performance. Mismatches were considered in all components of the main blocks (differential oscillatory neurons and synaptic circuits). First, they were applied to each block separately to study their effect. Sensitivity analysis was then used to find the components and parameters most critical to performance degradation. Simulation results showed that a mismatch in the components of differential oscillatory neurons, especially in the high threshold voltage of $VO_2$ devices, was more critical for performance degradation in terms of desynchronization and instability than a mismatch in

The authors are with the Instituto de Microelectrónica de Sevilla IMSE-CNM, Parque Tecnológico de la Cartuja, CSIC, 41092 Sevilla, Spain (e-mail: bernabe@imse-cnm.csic.es).

the synaptic connections. Finally, design space exploration was conducted to study DONN tolerance to mismatches when the *$VO_2$* parameters are varied.

The rest of the paper is organized as follows. Section 2 provides a brief introduction to ONNs, $VO_2$ devices, and memristors. Section 3 reviews DONN architecture, storing patterns, and retrieving patterns, and also introduces the method used to apply mismatches to the components. Section 4 presents the simulation results, sensitivity analyses, and design space exploration (DSE), and some conclusions are drawn in Section 5.

## II. Background

ONNs are dynamic systems made up of weakly connected oscillatory neurons. They are described through Ordinary Differential Equations (ODEs) ) [1], [17]

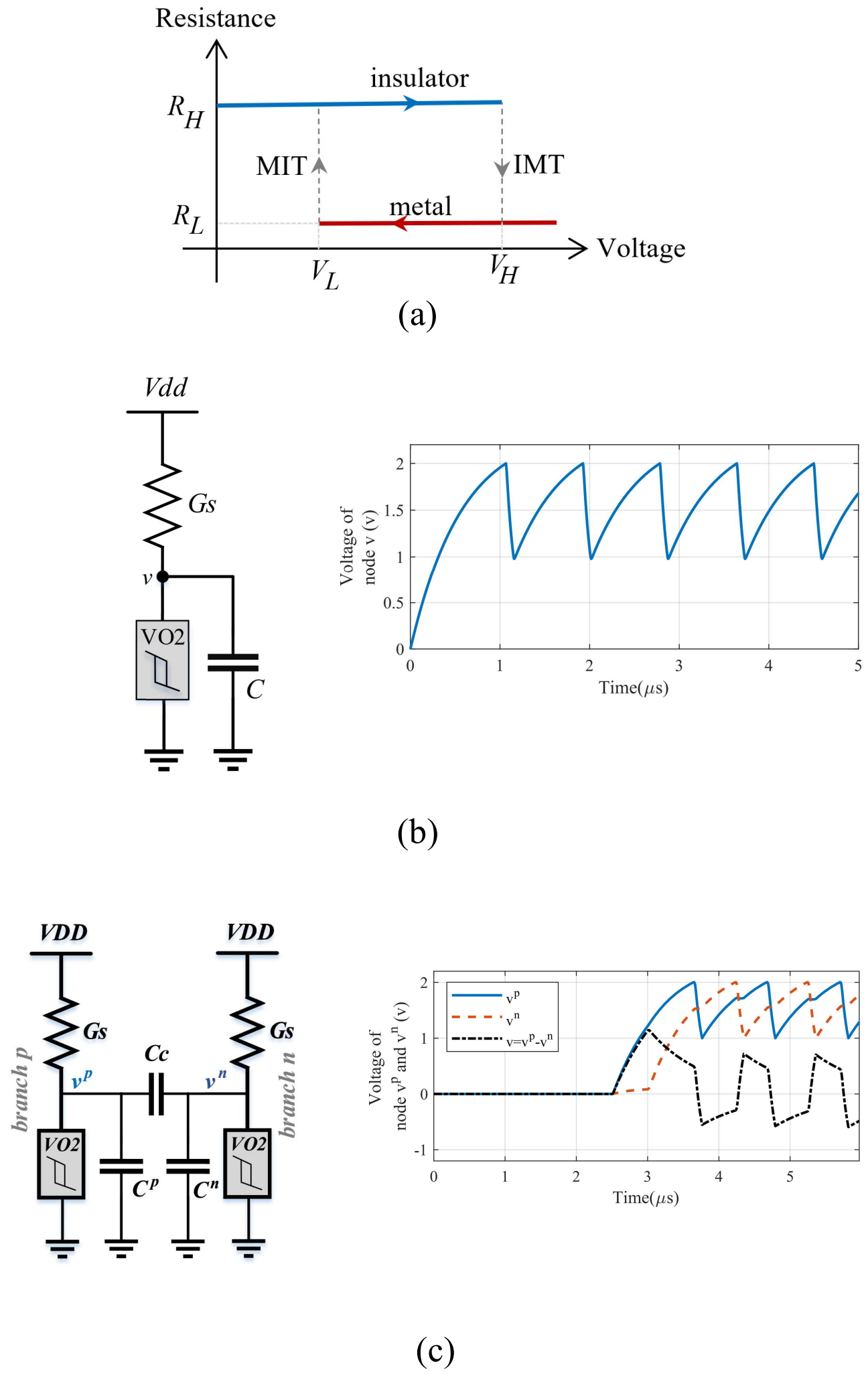

**Fig. 1.** (a) R-V characteristic of a a VO2 device. (b) A single-ended oscillator circuit and its output. (c) Circuit of the differential oscillatory neuron capable of producing differential signals. In these simulations, the parameters values are $R_S = 1/G_S = 6$ kΩ, $C_p$=$C_n$=$C$= 109 nF, $C_C$=10.9 nF, $V_{dd}$ = 2.5V, $V_H$ = 2V, $V_L$ = 1V, $R_L$ = 1kΩ, $R_H$ = 100kΩ, and $\tau$ = 100ns. The initial voltage of $C$= $C_p$=$C_n$ and $C_c$ is zero.

$$\dot{x}_i = f_i(x_i) + \sum_{j=1}^{n} w_{ij} g_{ij}(x_i, x_j)\,, i = 1,2,\dots,n \quad (1)$$

where $n$ is the number of the oscillatory neurons and $x_i$ is the state vector of oscillatory neuron $i$ ($x_i \in \mathbb{R}^m, m \geq 2$). Function $f_i(x_i)$ describes the dynamic behavior of oscillatory neuron $i$ and is usually formulated using m-dimensional differential equations. Parameter $w_{ij}$ is the adjustable weight between oscillatory neurons $i$ and $j$. Synaptic function $g_{ij}(x_i,x_j)$ defines the effect of oscillatory neuron $j$ on oscillatory neuron $i$. Broadly speaking, ONNs comprise two main blocks: oscillatory neurons $f_i(x_i)$ and adjustable synaptic functions $w_{ij}\ g_{ij}(x_i,x_j)$. In the following subsections, the hardware implementation of these two blocks is reviewed.

### A. Oscillatory neurons

Function $f_i(x_i)$ describes the dynamic behavior of oscillatory neurons that can be implemented using $VO_2$ devices. A $VO_2$ device is a two-terminal component based on a phase change material that presents insulator-to-metal (IMT) and metal-to-insulator (MIT) transitions [2]. The transitions are temperature-driven, caused by in-device joule heating in the presence of an applied voltage. Increasing the device temperature causes a change from a high resistance ($R_H$) state to a low resistance ($R_L$) state. Conversely, there is a transition from a low to a high resistance state when the device temperature decreases. From a circuit analysis point of view, this local in-device temperature dependence can be understood more simply as a simple terminal voltage dependence [2], [4], [8], [18]. In [18], a SPICE model for $VO_2$ devices is introduced where the transitions are related to a high threshold voltage $V_H$ and a low threshold voltage $V_L$. Fig.1(a) shows the R-V characteristic of a $VO_2$ device. When an increasing applied voltage reaches $V_H$, the resistance switches from its high resistance value $R_H = 1/G_L$ to its low resistance value $R_L = 1/G_H$. When a decreasing applied voltage drops below $V_L$, there is a transition from the low resistance state to the high resistance state. The time constant of the transitions is $\tau$. In this paper, simulations are based on the $VO_2$ model introduced in [18] (see the appendix for more details). Default values of the parameters are $V_H$ = 2V, $V_L$ = 1V, $R_L$ = 1kΩ, $R_H$ = 100kΩ, and $\tau$ = 100ns.

Fig.1(b) shows a single-ended oscillator comprising the series connection of a $VO_2$ device and a resistor with a capacitance load. Differential oscillatory neurons are designed by coupling two single-ended oscillators through a capacitor (Fig.1(c)). The coupling capacitor forces the positive branch p and negative branch n to stay in anti-phase, thus producing two signals of opposite phase (the time difference between the peak time of signal $v_p$ and $v_n$ is around half of the oscillators' period). The oscillator differential output is defined as $v^p$-$v^n$ and the oscillator period is determined by the following relation [5] which comprises two terms of rising time and falling time:

$$T = C^* \times \left[\left(\frac{1}{G_L + G_S}\right) ln(\frac{V_{max} - V_L}{V_{max} - V_H}) + \left(\frac{1}{G_H + G_S}\right) ln(\frac{V_{min} - V_H}{V_{min} - V_L})\right] \quad (2)$$

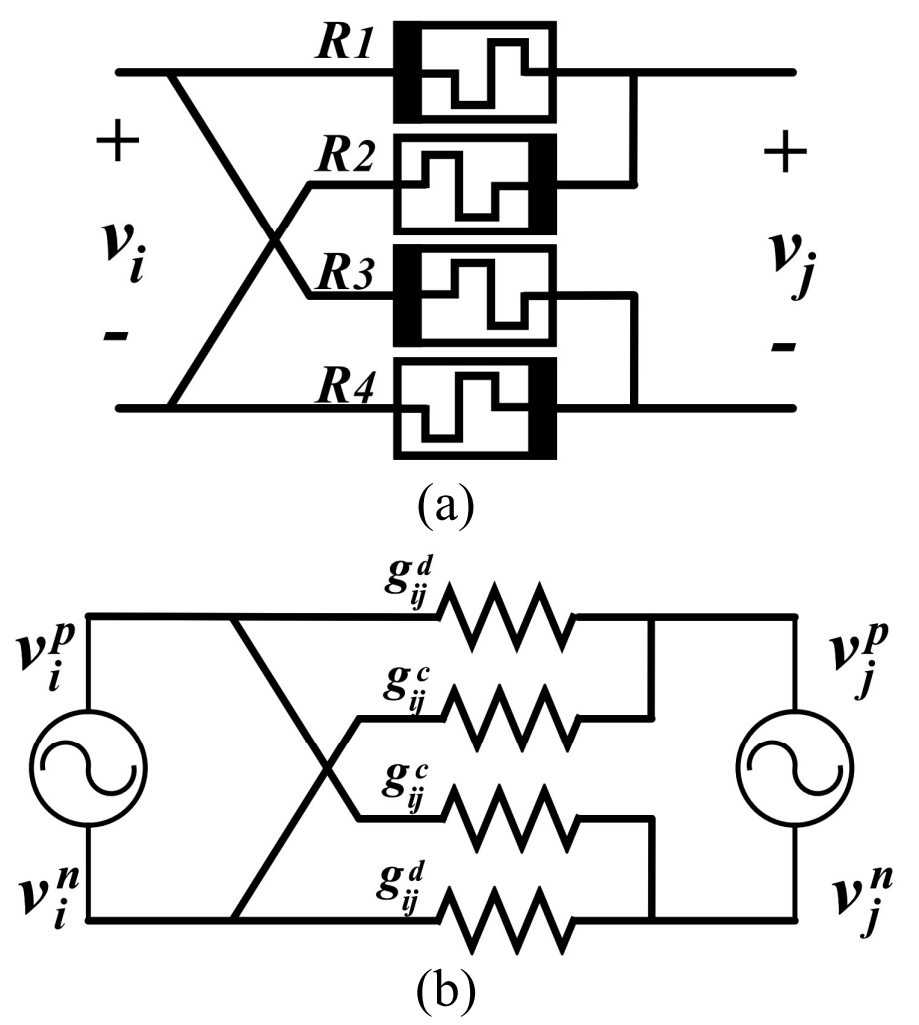

**Fig. 2.** (a) Memristor-bridge circuit. (b) Equivalent circuit of two differential oscillatory neurons coupled through a memristor-bridge circuit.

where $V_{max} = \frac{G_S Vdd}{G_L + G_S}$ and $V_{min} = \frac{G_S Vdd}{G_H + G_S}$. Parameter $C^*$ is the capacitance at the output nodes of the differential oscillatory neurons and is approximately considered as $C^* \approx C + C_c$. The capacitance of an output node is the capacitance of an equivalent capacitor between the oscillator output and ground.

### *B. Synaptic circuit*

Synaptic circuits can be designed using resistors to implement function $w_{ij}g_{ij}(x_i,x_j)$. Depending on the architecture of the oscillator circuit, they can be single-resistor circuits (compatible with single-ended oscillators) or Wheatstone-bridge circuits (compatible with differential oscillators). The resistors can be replaced by memristors to make the synaptic circuits adjustable. A memristor is a two-terminal resistive device whose resistance is adjustable. In addition to its application as an analog memory, memristor can be used in different applications such as machine learning [19] and dynamical systems [20], [21] . A memristor-bridge circuit is a counterpart of the Wheatstone-bridge circuit, which is compatible with differential oscillatory neurons, and can have positive, negative, or zero weights. Fig. 2(a) shows a memristor-bridge circuit in which the differential terminals allow differential oscillatory neurons to be connected. Fig. 2(b) is an equivalent memristor-bridge circuit coupling two differential oscillatory neurons. Conductance $g_{ij}^{(d)}$ is located between the positive branches (negative branches), tending to put positive branches (negative branches) in phase. When the positive branches (negative branches) are in phase, then the two differential oscillatory neurons are considered to be in phase. On the other hand, conductance $g_{ij}^{(c)}$ is located between a positive and a negative branch, thus tending to put them in phase while the positive branches (negative branches) will tend to be in anti-phase. When the positive branches (negative branches) are anti-phase, then the two differential oscillators are considered to be anti-phase. Depending on the relative values of $g_{ij}^{(d)}$ and $g_{ij}^{(c)}$, two coupled differential oscillatory neurons tend to be in phase or anti-phase. When $g_{ij}^{(d)} > g_{ij}^{(c)}$ ($g_{ij}^{(d)} < g_{ij}^{(c)}$) they tend to be in-phase (anti-phase). It is therefore possible to implement a coupling circuit with a positive (if $g_{ij}^{(d)} > g_{ij}^{(c)}$), negative (if $g_{ij}^{(d)} < g_{ij}^{(c)}$), or zero (if $g_{ij}^{(d)} = g_{ij}^{(c)}$) weight.

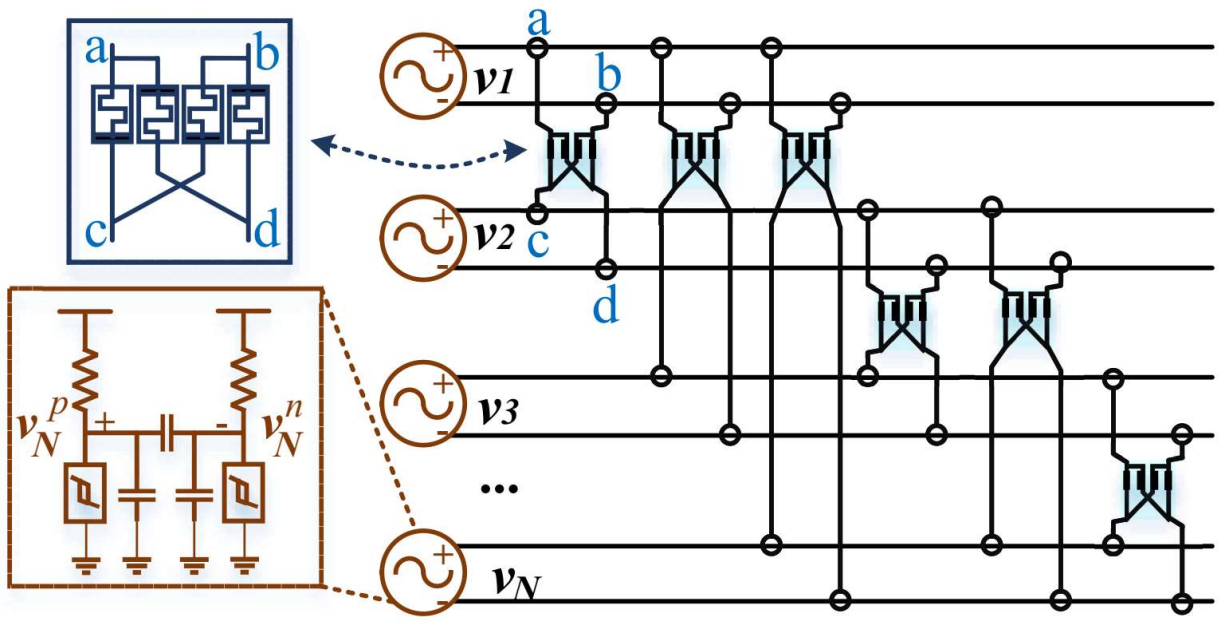

**Fig. 3.** Hardware architecture of the fully connected DONN. Differential neurons are implemented with a pair of $VO_2$ oscillators, oscillating in anti-phase. Synapses are implemented using memristor-bridge circuits.

## III. Differential ONNs (DONNs)

DONNs comprise differential oscillatory neurons and memristor-bridge circuits, as shown in Fig. 3. Their fully connected architecture is an oscillatory counterpart of the Hopfield Neural Network [22] which can be used as an associative memory to store and retrieve patterns.

### *A. Storing patterns*

Here, a learning rule was used to store patterns, adjusting the synaptic weights accordingly. A pattern was represented by a vector with $N$ elements (pixels). Only binary values were considered. Once the weights were known, a mapping rule was used to obtain the physical resistances for the memristor-bridge synapses. To store patterns in a DONN, we used the following Hebbian rule to calculate the weights [23]

$$w_{ij} = \frac{1}{N}\sum_{k=1}^{P} b_i^{(k)} b_j^{(k)} \quad i,j \in \{1,2,3,\ldots,L\}, b_{i,j}^{(k)} \in \{-1,+1\} \tag{3}$$

where $P$ was the number of stored pattern, $N$ was the number of pixels in each pattern (equal to the number of neurons in the DONN), and $b_i^{(k)}$ is the binary element $i$ of pattern $k$. Elements $b_i^{(k)}$ and $b_j^{(k)}$ of all stored patterns were used to calculate weight $w_{ij}$. Using the Hebbian rule, the weight between the neurons that were corresponding to the pixels with the same value was increased. In other words, for a given pattern, if pixel $i$ and $j$ had the same value (either -1 or +1), the weight between neuron $i$ and $j$ would be increased. On the other hand, if the value of pixel $i$ and $j$ was different, the weight between the corresponding neurons decreased.

The following rules were used to map the sign and value of the above weights to the memristors' resistances. Weights $w_{ij}$ were mapped to the $g_{ij}$ values using the following relation [5]

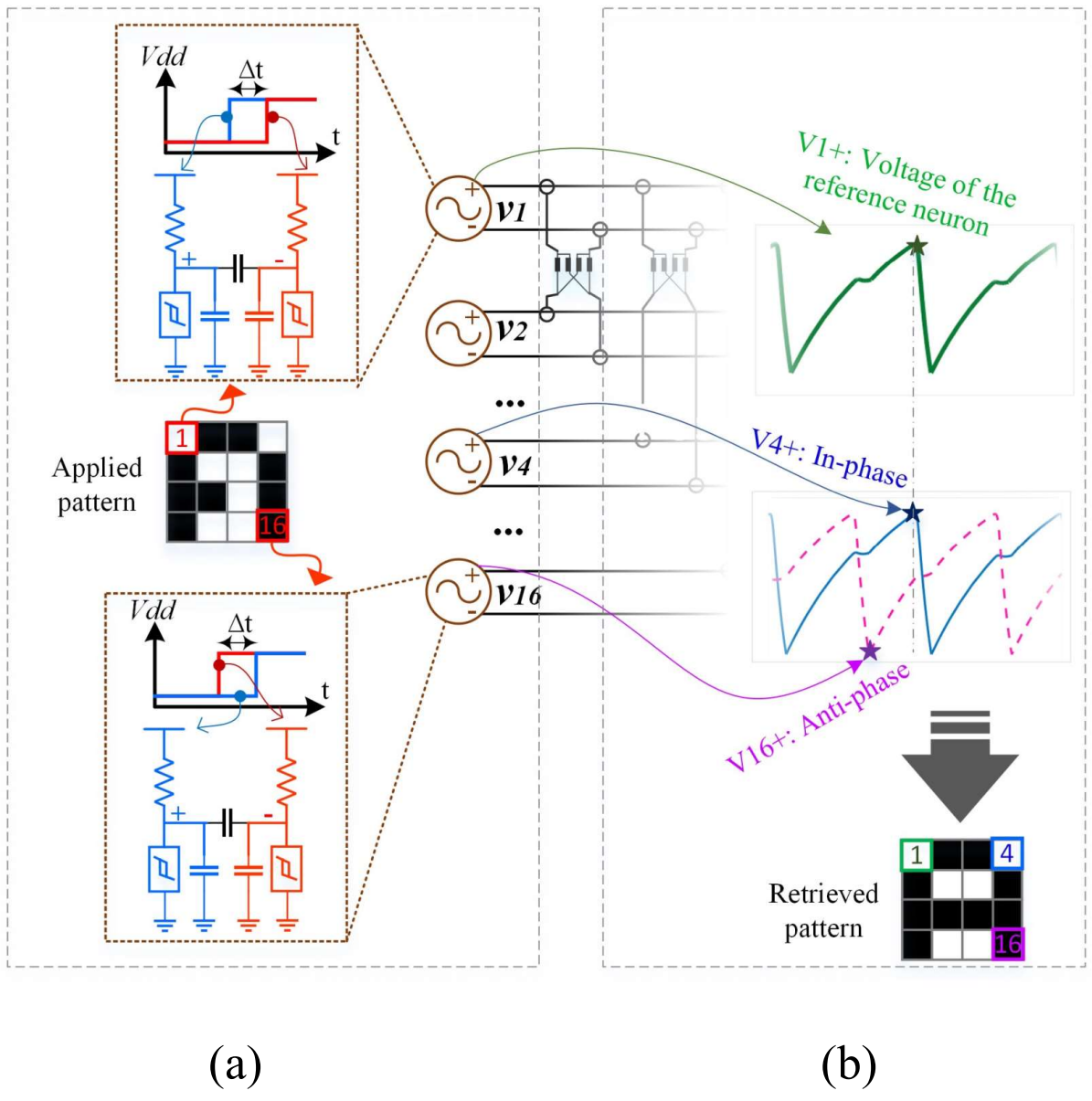

**Fig. 4.** (a) Application of an input pattern. For white pixels, the power supply is applied first to the positive branch of the corresponding neuron and then, after a one-half period, to the negative branch. (b) After convergence, the phase of the positive branches is compared with the phase of the reference neuron to determine which neurons are in-phase or anti-phase, corresponding to the white and black pixels of the retrieved pattern, respectively.

$$g_{ij} = \begin{cases} \dfrac{g_0}{1+\beta \times P \times |m_{ij}|_{norm}} & ,w_{ij} \neq 0 \\ \dfrac{g_0}{1+\beta \times P} & ,w_{ij} = 0 \end{cases} \quad (4)$$

where parameter $\beta$ was a small positive value (e.g., 0.2) that controlled the mapping range for conductance $g_{ij}$. Value $|m_{ij}|_{norm}$ was the Min-Max normalization of the matrix $M = (|\frac{1}{w_{ij}}|) \in \mathbb{R}^{L\times L}$. To obtain the Min-Max normalization, the following relation is used

$$\left|\frac{1}{w_{ij}}\right|_{norm} = \frac{\left|\frac{1}{w_{ij}}\right| - min}{max - min} \quad (5)$$

where $min$ and $max$ are the minimum and maximum values of $|1/w_{ij}|$ among all non-zero weights, respectively. A larger value for $\beta$ provided a larger difference between the values of conductance $g_{ij}$. In other words, for a given number of patterns $P$, there were discrete values of weights $w_{ij}$ and $|m_{ij}|_{norm}$. By increasing $\beta$, the difference between the discrete results of the denominator in (4) was increased, and consequently the difference between the different values of conductance $g_{ij}$ was increased.
The following mapping rule was also used to map the weight signs to resistance values $g_{ij}^{(d)}$ and $g_{ij}^{(c)}$

$$g_{ij} = \begin{cases} \alpha g_{ij}^{(d)} = g_{ij}^{(c)} & w_{ij} < 0 \\ g_{ij}^{(d)} = \alpha g_{ij}^{(c)} & w_{ij} > 0 \\ \alpha g_{ij}^{(d)} = \alpha g_{ij}^{(c)} & w_{ij} = 0 \end{cases} \quad (6)$$

where $\alpha > 1$ was a constant value. The default value of $\alpha$ is 1.8 for all simulations.

Parameter $g_0$ is the maximum conductance (inverse of the minimum resistance) of memristors $g$, and was determined using the following conditions [5]

$$\frac{G_s V_{dd} + g(N-1)V_L}{(G_s + G_L + (N-1)g)} > V_H \quad (7)$$

$$\frac{G_s V_{dd} + g(N-1)V_H}{(G_s + G_H + (N-1)g)} < V_L \quad (8)$$

where $N$ was the number of differential oscillatory neurons. These conditions guaranteed that the voltage of the oscillators would not reach stable points, resulting in permanent oscillation [24]. For more details see [5].

### *B. Applying test patterns and retrieving stored patterns*

Input patterns were applied to DONNs via phase initialization of the differential oscillatory neurons. Each pixel of the input pattern was applied to the corresponding differential oscillatory neuron. Depending on the binary pixel value, the power supply was applied first to one of the branches of the differential oscillatory neuron, and then, after a specific delay, to the second branch. The applied delay time had to be half of the oscillators' period. Fig. 4(a) shows an example of applying the first and last pixels of a pattern to the corresponding differential oscillatory neurons. The first pixel is white (equivalent to binary value 1) and the power supply was applied first to the positive branch of the corresponding differential oscillatory neuron and then, after a one-half period, to the negative branch. In contrast, for the black pixel (equivalent to binary value -1), the positive branch was powered up after the negative branch.

When an input pattern is applied to a trained DONN, differential oscillatory neurons start to change their phases and synchronize with each other either in-phase or anti-phase, depending on the stored patterns. By identifying the phase of the neurons with respect to a reference neuron (the first neuron), the retrieved pattern is determined. For a given neuron that is in-phase (anti-phase) with the reference neuron, a white (black) pixel value is assigned. In this study, a pattern and its complement were interchangeable. Fig. 4(b) shows the retrieval of a stored pattern in which the positive voltages were compared with the reference neuron to determine whether the neurons were in-phase or anti-phase.

### *C. Evaluation*

To evaluate DONN performance, three specifications were considered: retrieval accuracy, synchronization level, and stability rate. Retrieval accuracy was the number of correct retrieved patterns against the total number of applied patterns. Synchronization level was defined to measure the degree to which the differential oscillatory neurons were synchronized either in phase or in anti-phase. For an input pattern $p$, the synchronization level in cycle $c$ was a value in the range of [0 1] that showed how much the neurons were synchronized with each other, either in-phase or anti-phase. When the neurons were not exactly in-phase or anti-phase, a value less than 1 was assigned, depending on the phase difference. For an input pattern $p$, the synchronization level in cycle $c$ was calculated using the following relation [5]

$$SYN_p(c) = <\frac{1}{N}\sum_{i=1}^{N} m(PT_i(c) - PT_{ref}(c))> \quad (9)$$

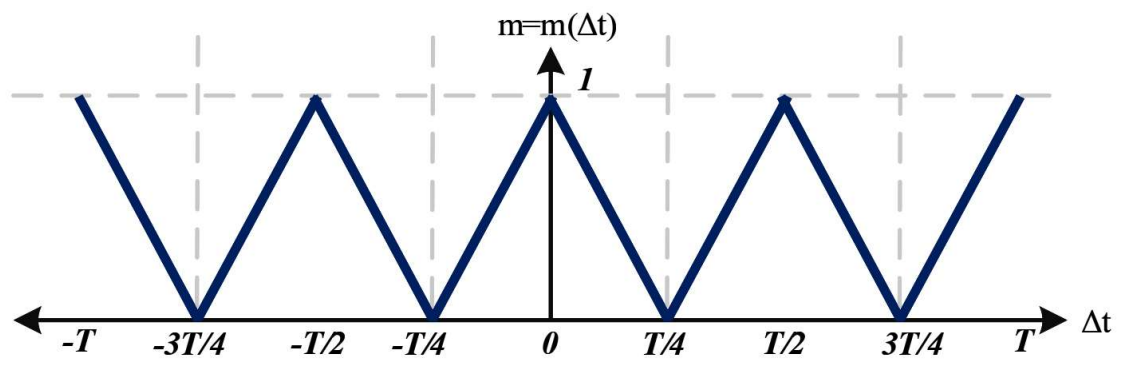

**Fig. 5.** Function $m=m(\Delta t)$ used to map a time difference to a value between 0 and 1

The $SYN_p(c)$ value shows the synchronization level at cycle $c$, which is related to the time difference between the peak time of signal $i$ at cycle $c$, $PT_i(c)$, and the peak time of the reference signal $PT_{ref}(c)$ (the signal of positive branches $v_i^p$ was used for calculations). Function $m=m(\Delta t)$ maps a time difference $\Delta t$ to a value between 0 to 1 (Fig. 5.).

Stability rate was defined as a value in the range of [0 1] corresponding to the number of applied patterns resulting in stable outputs divided by all the applied patterns.

$$STB = \frac{\#P_{stable}}{\#P_{applied}} \quad (10)$$

where $\#P_{stable}$ is the number of patterns that result in a stable output and $\#P_{applied}$ is the number of all applied patterns.

The output was stable if, after a pattern retrieval, the retrieved pattern did not change (i.e., the phase of the neurons did not change with respect to each other). On the other hand, for an unstable output, the extracted pattern changes over time.

### *D. Applying mismatches*

Mismatches between the parameters of identical devices are the result of random processes that occur during the fabrication phase. In this paper, mismatches were considered in the parameters of all devices and DONN performance was evaluated in the presence of those mismatches. In this regard, for a given parameter $x$, samples with normal distribution were generated by

$$\tilde{x} \sim N(x, \sigma) \quad (11)$$

where $\tilde{x}$ is a set of samples that are normally distributed with mean $x$ and variance $\sigma^2$. For a given parameter, the number of samples equaled the number of the corresponding devices in the DONN's architecture. The device parameters are listed in Table I. They are categorized into parameters of differential oscillatory neurons and synaptic circuits.

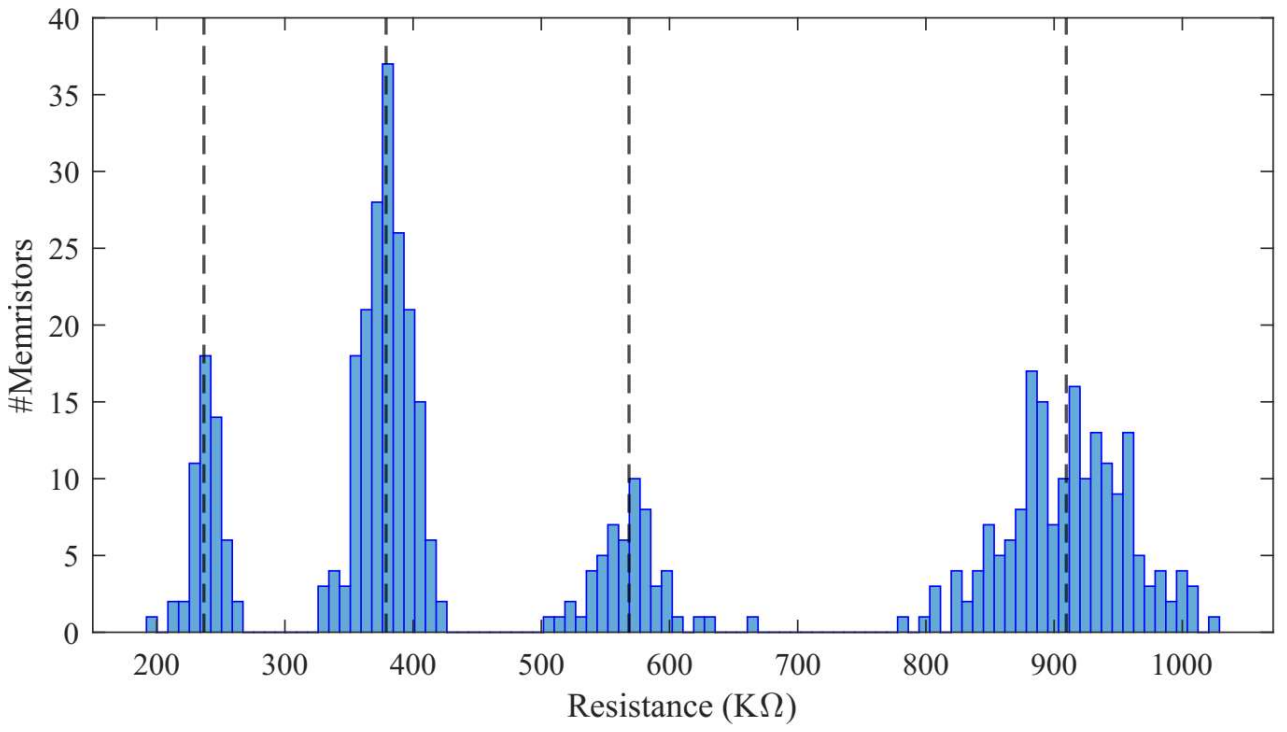

**Fig. 6.** Normal distribution (RSD= 5%) of the memristor resistance in a DONN with $N$=16 neurons and $P$=3 stored patterns.

## IV. Results

This section describes the simulation of the DONNs in a SPICE simulator, with mismatches in the main blocks (synaptic circuits and differential oscillatory neurons). The effects of mismatches in each block were studied separately to evaluate the DONN's tolerance. Sensitivity analysis was then used to obtain the components and most critical parameters for performance degradation.

### *A. Mismatches in the synaptic circuits*

In this part of the study, mismatches were considered in the synaptic circuits, caused by deviations in the resistance of memristors. It was assumed that memristors were already programmed to the desired resistance and they were treated as resistors with mismatches. Random resistance values with normal distribution were generated using Eq. (11), in which the mean value $\mu$ was the nominal value of a given resistance. As an example, Fig. 6 shows the resistance histogram for the synaptic circuits of a DONN when the resistance relative standard deviation (RSD) $\sigma$ was 5%. In this case, the number of neurons was $N$=16 and the number of stored patterns was $P$=3. Note that for $N$=16, the number of memristors was $2N.(N-1)$=480. In addition, only four different nominal memristor

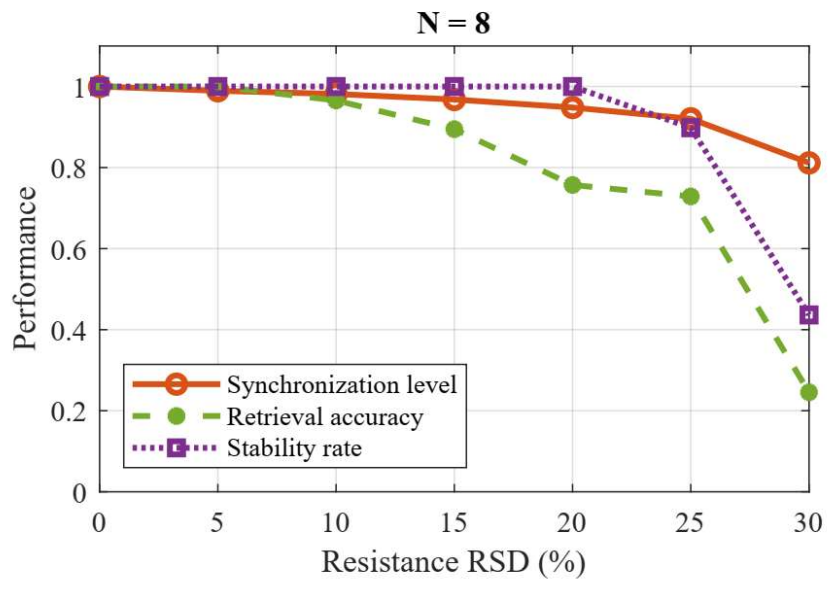

(a)

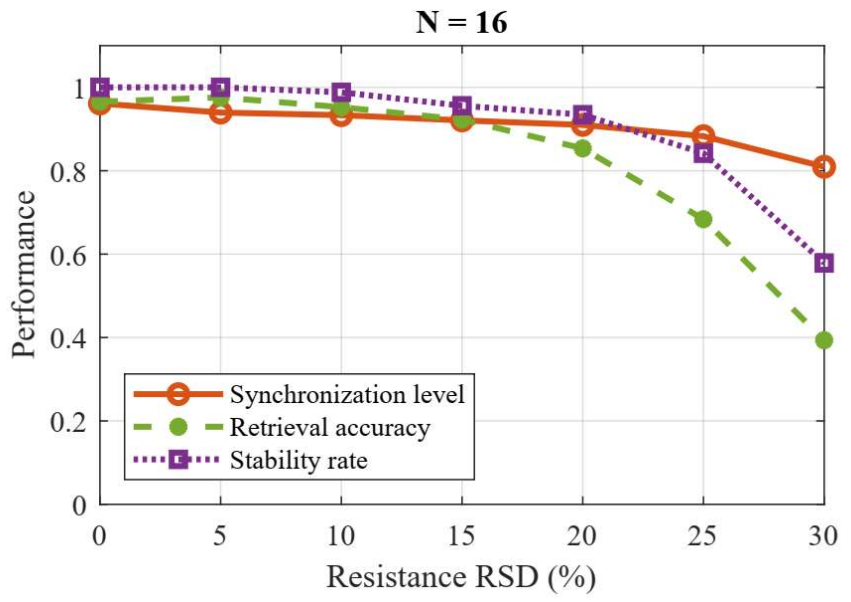

(b)

**Fig. 7.** DONN performance ($N$=8, 16) in the presence of mismatches in the memristors of the synaptic circuits. (a)-(b) The DONN's synchronization level, retrieval accuracy, and stability rate change according to the resistance RSD.

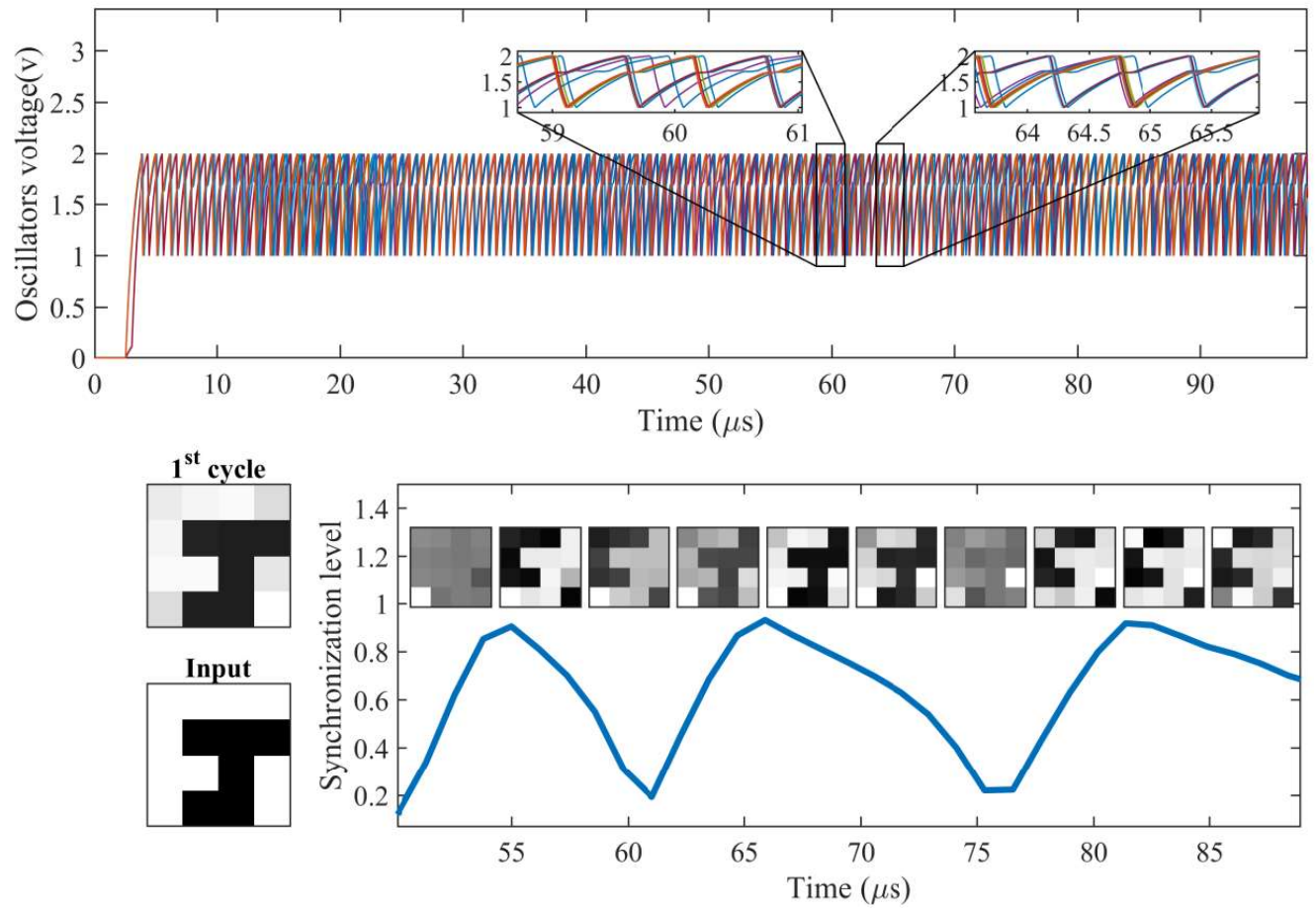

**Fig. 8.** Simulation of a DONN with $N$=16 with a resistance RSD of 30%, showing unstable behavior.

resistance values were obtained using the Hebbian rule and mapping rule (see Fig. 6).

To evaluate DONN performance ($N$=8, 16), independent simulations were performed 10 times for each resistance RSD value (for each simulation, mismatches are different). Other DONN parameters are summarized in Table I. Fig. 7(a)-(b) shows DONN performance ($N$=8, 16) in terms of synchronization level, retrieval accuracy, and stability rate for different resistance RSDs. Synchronization level and stability were both above 90% when RSD was less than 20%. Retrieval accuracy was also robust up to 15% mismatches but then fell as RSD continued to increase. These results show that differential oscillatory neurons can synchronize in the presence of relatively large mismatches in synaptic weights (up to 20%). Instability arose when the mismatches were further increased. Fig. 8 shows an unstable operation when the resistance RSD was 30%. The top figure shows the voltages of the positive branches and the bottom figure includes the input pattern, the 1st cycle pattern, the synchronization level, and the evolution of the retrieved pattern. The output pattern did not converge to a specific pattern but altered continuously.

TABLE I. PARAMETERS OF DONNS WITH FREQUENCY OF f=1/T=1MHZ

| Parameters | | Value |
|---|---|---|
| Number of neurons | $N$ | 8, 16 |
| Supply voltage | $Vdd$ | 2.5 V |
| $VO_2$ parameters | $V_H$ | 2 V |
| | $V_L$ | 1 V |
| | $R_H=1/G_L$ | 100kΩ |
| | $R_L=1/G_H$ | 1kΩ |
| Series resistance | $Rs=1/Gs$ | 6kΩ |
| Mapping parameters | $\alpha$ | 1.8 |
| | $\beta$ | 0.2 |
| Coupling capacitor | $C_C$ | 10.9 pF |
| Parallel capacitor | $C$ | 109 pF |

*B. Mismatches in the differential oscillatory neurons*

In general, mismatches in natural frequencies prevent oscillators from synchronizing. For instance, two coupled Kuramoto oscillators can only tolerate limited mismatches of natural frequencies dependent on the coupling strength [25]. The same constraint also exists for DONNs. Differential oscillatory neurons comprise $VO_2$ devices, resistors, and capacitors, and mismatches change the specifications of the neurons, specifically their natural frequency. Considering Eq. (2), the natural frequency of differential oscillatory neurons is related to the main component parameters, which are listed in Table. II. Variations of each parameter directly cause variation in the natural frequency of neurons. Non-uniformity of frequencies can prevent oscillators from synchronizing, and consequently cause instability in DONNs.

This work studied the effect on DONN performance of mismatches in the components of differential oscillatory neurons. To do so, DONNs with different numbers of neurons ($N$=6, 8, 12, 14, 16) were designed. For a given size of DONN ($N$), random patterns were generated and stored in DONNs where the number of stored patterns was 2 for $N$=6, 8, and 3 for $N$=12, 14, 16. Random test patterns were also generated for evaluation (the number of test patterns was 1.5$N$). Mismatches were then applied to the component parameters separately. The seven parameters shown in Table II correspond to the differential oscillatory neurons. For a given parameter, Eq. (11) was used to generate $M$ random values with normal distributions where $M$ was the number of components used in a DONN with $N$ neurons (number of components are: #Memristor = 2N(N-1), #Capacitor_C = 2N, #Capacitor_$C_C$ = N, #Resistor = 2N, #$VO_2$ = 2N). The nominal value of the parameter was used as the mean value $\mu$, and the standard deviation $\sigma$ was increased to evaluate the tolerance of DONNs to the component mismatches. Simulations were performed with the aforementioned variables ($N$ and component RSD) and test patterns were used for evaluation. Finally, the average results were calculated for each given component *RSD* by applying the test patterns to the network of different size $N$. The results are shown in Fig. 9. Table II maps the RSD ranges in Fig. 9 to the corresponding ASD (absolute standard deviations) used in this study. In each figure, the top horizontal axis shows a parameter RSD and the bottom axis is the natural frequency RSD. Each data point is the average of simulations with different number of neurons ($N$=6, 8, 12, 14, 16) and random patterns (1.5$N$) for a specific RSD value. The results show that the DONN's performance, in terms of synchronization level, retrieval accuracy, and stability rate, decreased when the parameter RSD and, equivalently, the natural frequency RSD increased. The last figure is the average of all the results, showing the performance of the DONN with respect to the natural frequency RSD. Here, performance dropped dramatically when the natural frequency RSD was above 0.5%, showing that DONNs are more vulnerable to neuron mismatches than to synaptic circuit mismatches. The overall effect of mismatches in differential oscillatory neurons could also be considered as frequency deviation, with performance degradation being caused by the non-uniformity of frequencies. It was therefore worth subjecting the natural frequency of neurons to sensitivity analysis to determine the most critical parameters. The sensitivity analysis is described in the next subsection.

TABLE II. MAPPING BETWEEN RSD AND ASD RANGES IN FIG. 9.

| | **nominal** | **$RSD_{min}$** | **$RSD_{max}$** | **$ASD_{min}$** | **$ASD_{max}$** |
|---|---|---|---|---|---|
| Natural freq. | 1MHz | 0% | 0.6% | 0 | 6KHz |
| $V_H$ | 2V | 0.03% | 0.17% | 0.6mV | 3.4mV |
| $V_L$ | 1V | 0.13% | 1% | 0.13mV | 9.7mV |
| $R_H$ | 100kΩ | 0.8% | 5% | 0.8kΩ | 5kΩ |
| $R_L$ | 1kΩ | 1% | 5.3% | 10Ω | 53Ω |
| $R_s$ | 6kΩ | 0.1% | 0.7% | 6Ω | 42Ω |
| $C$ | 109pF | 0.12% | 1% | 0.13pF | 1.1pF |
| $C_C$ | 10.9pF | 1% | 7.5% | 0.11pF | 0.82pF |

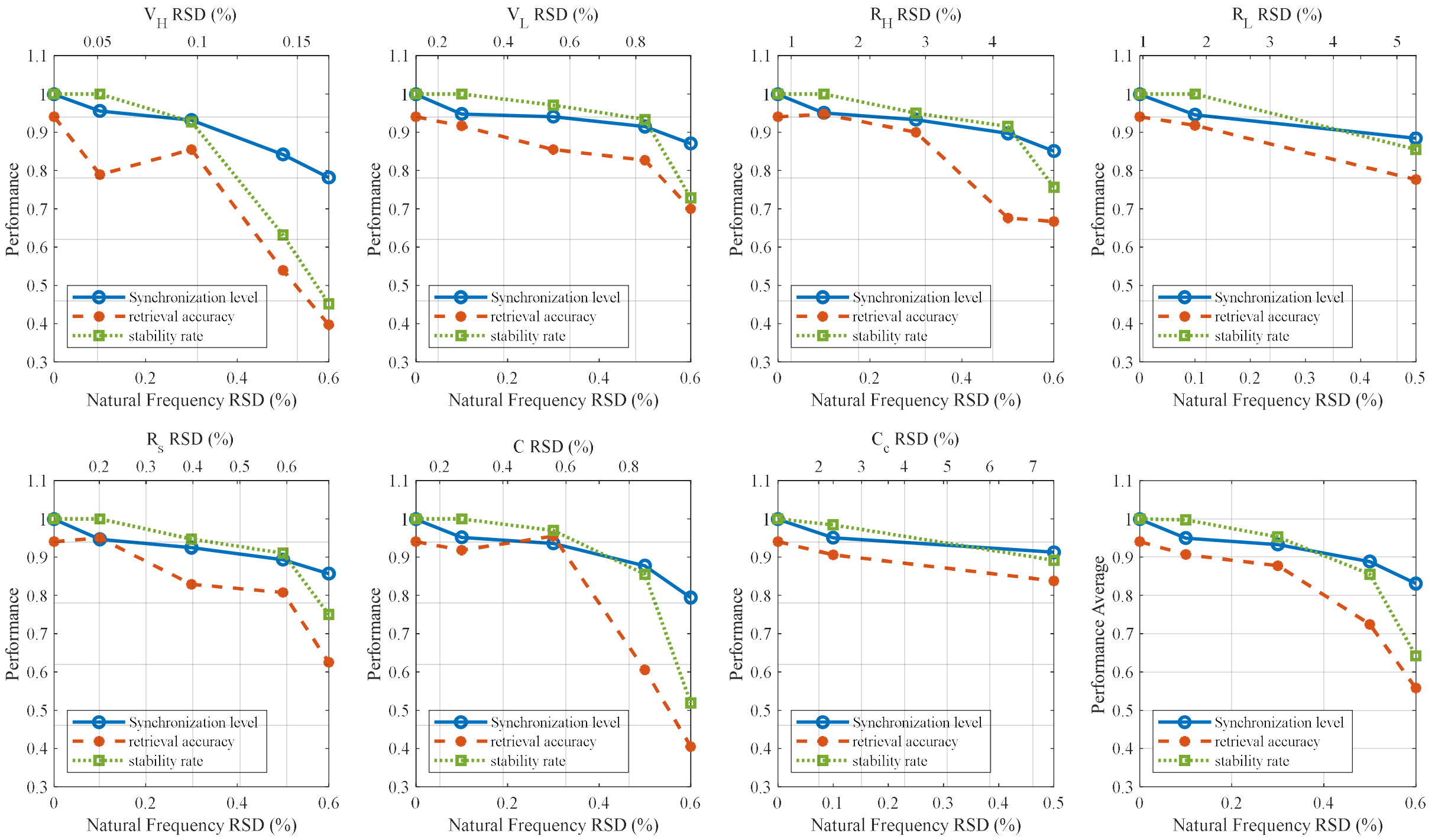

**Fig. 9.** Simulation results when the parameter RSD increases. In each plot, the top axis is the parameter RSD and the bottom axis is the natural frequency RSD. Performance is shown in terms of synchronization level, stability rate, and retrieval accuracy. The last figure is the average of all the figures.

C. Sensitivity analysis

Fig. 9 shows that performance degradation is related to natural frequency RSD. Natural frequency RSD also depends on parameter RSD. One of the differences between the parameters is how they impact the natural frequency RSD. Some parameters, such as $V_H$, have a stronger impact on the natural frequency RSD, while others have less impact. With this in mind, sensitivity analysis was performed to determine the impact of each parameter variation on the natural frequency variation. One approach to sensitivity analysis is to take the partial derivative of the output (frequency $f$) with respect to an input (parameter $x$)

$$S_x^f = \frac{x}{f}\frac{\partial f}{\partial x} \quad (12)$$

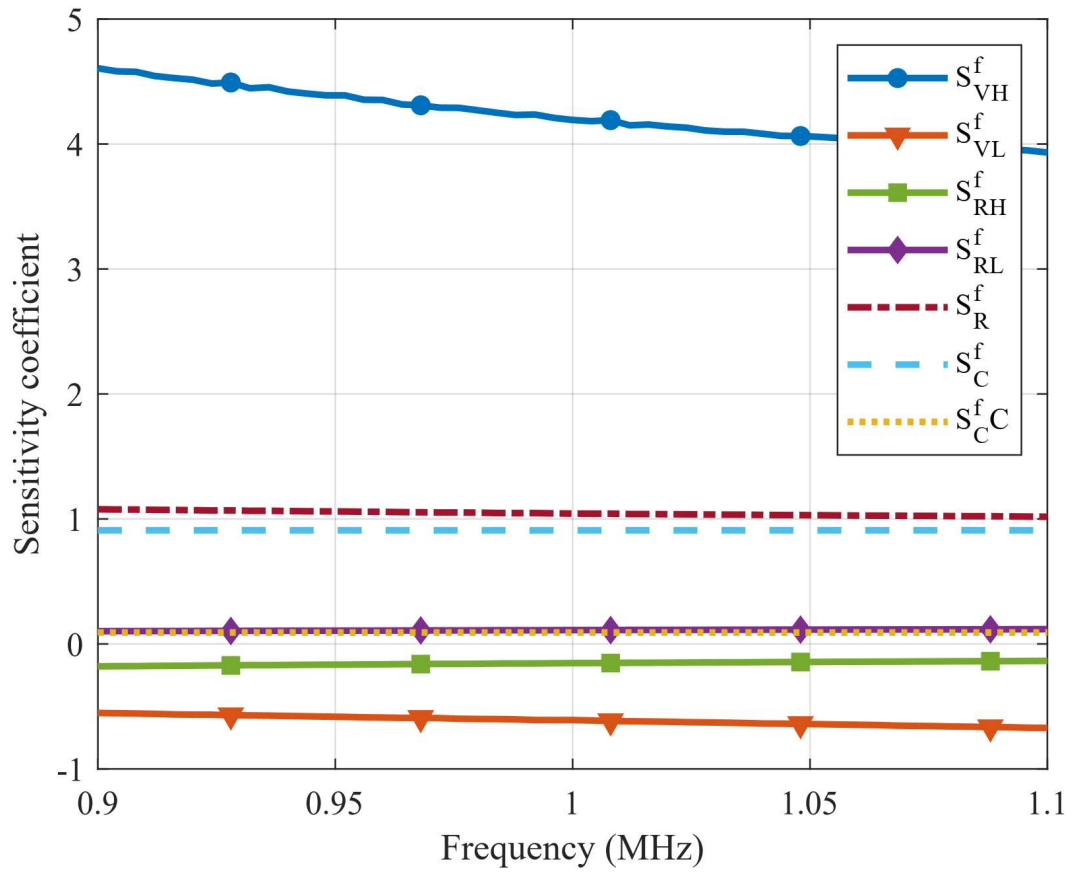

(a)

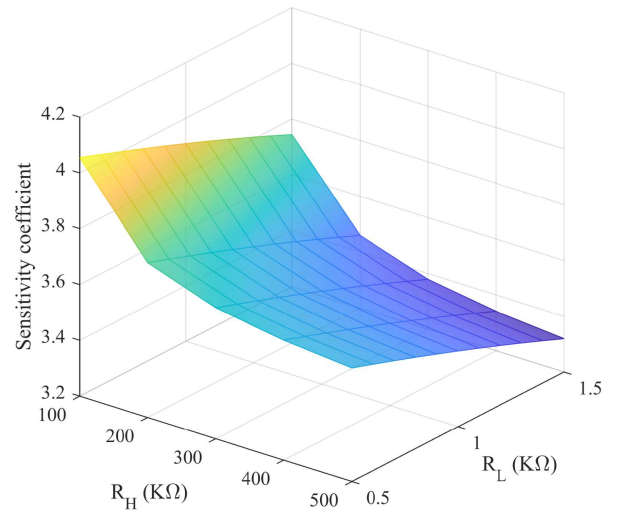

(b)

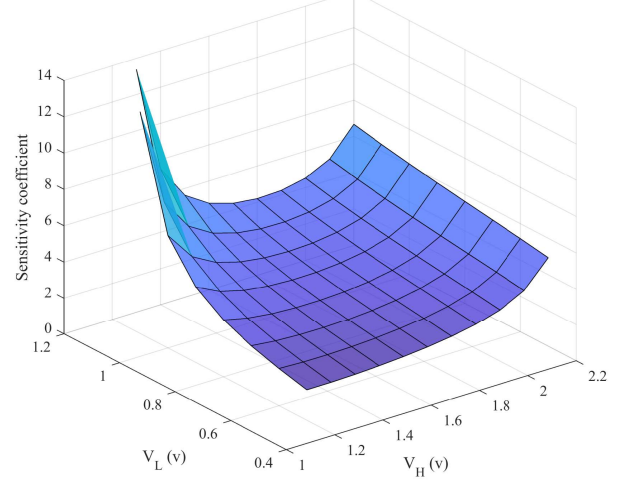

(c)

**Fig. 10.** Sensitivity of the natural frequency to the component parameters (a). Dependence of $S_{VH}^f$ to the parameters of the $VO_2$ device in planes $R_H$-$R_L$ (b) and $V_H$-$V_L$ (c).

where $x$ and $f$ are the nominal value of a parameter and the natural frequency of the isolated oscillators, respectively. The ratio $x/f$ was introduced to normalize the sensitivity coefficient $S_x^f$.

Fig. 10 (a) shows the sensitivity of the natural frequency with respect to the component parameters, calculated using Eq. (2) and Eq. (12) around the nominal frequency $f$=1MHz. The sensitivity analysis showed that the most critical parameter was the high threshold voltage of the $VO_2$ devices ($V_H$), while the low resistance $R_L$ and the coupling capacitance $C_C$ had the least impact on frequency variation.

Sensitivity coefficient $S_{V_H}^f$ was also calculated when the main parameters of $VO_2$ devices were altered. Figs. 10(b)-(c) show $S_{V_H}^f$ in planes $R_H$-$R_L$ and $V_H$-$V_L$, respectively. As can be seen in Figs. 10(a)-(b), $S_{V_H}^f$ was lower for larger $R_L$ and $R_H$ and lower $V_L$ and $V_H$.

### D. *Mismatches in single-ended ONNs*

In this subsection, a conventional ONN [26] based on the single-ended oscillatory neuron is studied (Fig. 11). The parameter values of the single-ended oscillatory neuron were the same as the nominal values of the differential oscillatory neurons listed in Table. II. The synaptic circuit was comprised of a memristor in parallel with a capacitor. The capacitor had a fixed value while the synaptic weights were mapped to the resistance of the memristor.

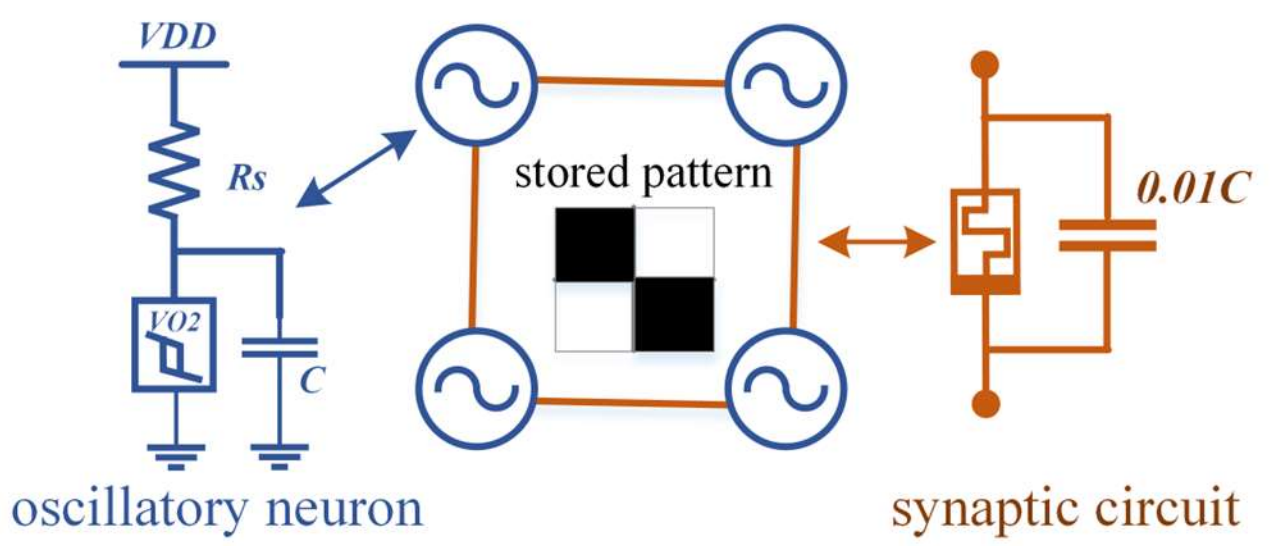

**Fig. 11.** A conventional ONN with four single ended oscillatory neurons.

The target ONN included four neurons and one pattern was stored in it. The synaptic weights were calculated using the Hebbian rule, then each weight was mapped into the values of the memristor and capacitor in the corresponding synaptic circuit. Due to the storage of one pattern, there were two distinct values for the synaptic weights (a negative and a positive value). The value of the capacitor in the synaptic circuit was 0.01C for both negative and positive weights. The resistance of the memristor was 8kΩ and 180kΩ for the positive and negative synaptic weights, respectively. To compare the results, a DONN counterpart with four neurons was simulated, accordingly.

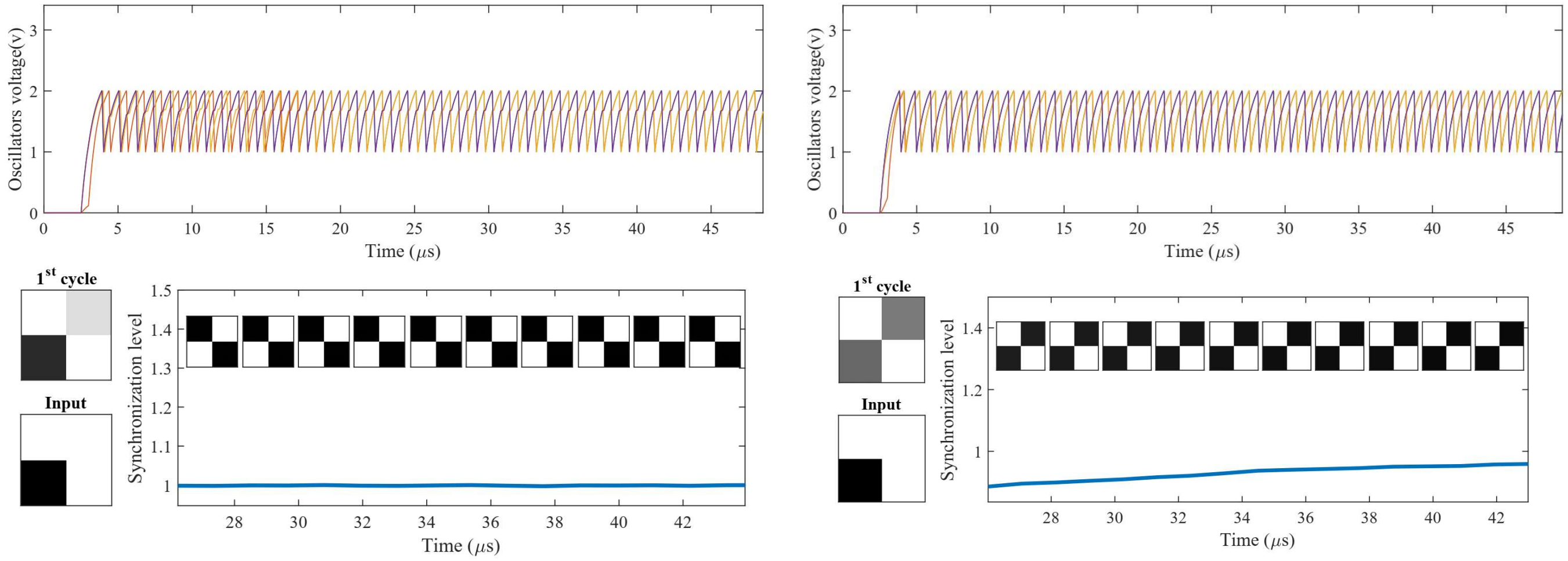

(a) (b)

**Fig. 12.** Simulation of ONNs with differential oscillatory neuron (a) and single-ended neuron (b). The input pattern and stored pattern were same for both cases and the output is retrieved correctly for both cases.

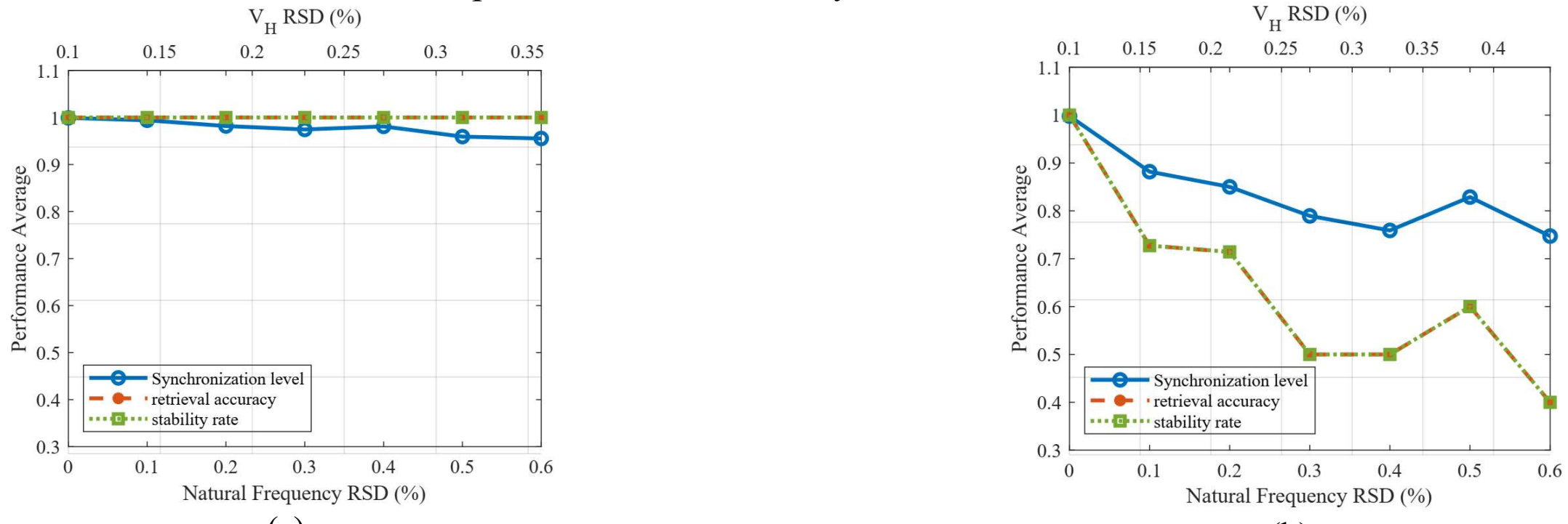

(a) (b)

**Fig. 13.** Simulation results of ONNs with differential oscillatory neuron (a) and single-ended neuron (b) in presence of mismatches in $V_H$. The horizontal axes in top and bottom of the figures are $V_H$ and corresponding natural frequency RSD, respectively.

Fig. 12 shows the simulation results when a random input is applied to the ONN and DONN. In this simulation, mismatches were not applied to the components. In both networks, the stored pattern was retrieved correctly.

To show the effect of the mismatches, random variations were applied to parameter $V_H$ as a critical parameter using (11). $RSD_{VH}$ (RSD of $V_H$) was changed from 0.001 to 0.005 and 10 independent simulations were performed for each value of $RSD_{VH}$. Fig. 13(a)-(b) shows simulation results for the DONN and single-ended ONN. The bottom horizontal axis is the natural frequency RSD corresponding to the $RSD_{VH}$ (in the figures the range of the natural frequency RSD is shown from 0 to 0.6%). The results show that the single-ended ONN is much more vulnerable to mismatches in comparison to DONN. Although in this case, the DONN was more tolerant to the mismatches compared to the single-ended ONN, its performance dropped dramatically for the larger size of the network was increased (see Section. IV (B)).

## E. *Design space exploration*

Design Space Exploration (DSE) is the methodical examination and elimination of undesirable design points based on relevant criteria. During DSE, variety of different design parameters can be explored, such as effective parameters of devices. Multiple optimization objectives, such as performance, power consumption, and cost, are considered simultaneously.

In this paper, the main parameters of $VO_2$ devices were explored. For this purpose, parameters $V_H$, $V_L$, $R_H$, and $R_L$ were varied. The other parameters had default values and are summarized in Table I. The performance in terms of synchronization level, stability rate, and retrieval accuracy was considered as optimization objective. Performance was optimized in presence of mismatches of parameter $V_H$ as a critical parameter.

For simplicity, DSE was conducted in two steps. First, parameters $V_H$ and $V_L$ were varied and the best values for them were selected to give maximum performance in terms of synchronization level, stability, and accuracy. Then parameters $R_H$ and $R_L$ were altered and performance was evaluated. The number of neurons was then $N$= 8, 16 and the number of the stored patterns was 2 for $N$= 8 and 3 for $N$=16. Random test patterns were also generated for evaluation (the number of test patterns was 1.5$N$).

Figs. 11 (a-f) illustrates DONN performance when parameters $V_H$, $V_L$, were varied. $V_H$ was made larger than $V_L$ ($V_H \geq V_L$+0.4V) and $RSD_{VH}$ (RSD of $V_H$) was changed from 0.001 to 0.005 (three values are shown in the figures). Fig. 14 (a) and Fig. 14 (d) show the synchronization levels for $N$=8 and $N$=16, respectively. When $RSD_{VH}$ increased, the synchronization level decreased accordingly. The synchronization level also dropped when $V_H$ increased. It is worth to mention that $V_{dd}$ is constant and equal to 2.5V. Similar results were also observed for stability (Fig. 14(b) and Fig. 14(e) ). However, accuracy (Fig. 14(c) and Fig. 14(f)) was low for small $V_H$ and $V_L$ values. It was observed that oscillatory neurons usually stuck at their initial state and their phases were not changed when $V_H$ and $V_L$ values are small.

To select the best values for $V_H$ and $V_L$, the maximum value for $RSD_{VH}$ at each point was obtained in which the synchronization level, stability, and accuracy were larger than 80%.

Fig. 14 (h) ($N$=8) and Fig. 14 (i) ($N$=16) show the maximum $RSD_{VH}$ values at each point that satisfied the abovementioned condition. Considering these figures, $V_H$ and $V_L$ would have to be selected from those values corresponding to the maximum $RSD_{VH}$ to obtain the highest level of robustness to $V_H$ variations. In Fig. 14(h), the optimum point was obtained with $V_H$=1.4 and $V_L$=0.6, allowing a maximum $RSD_{VH}$ of 0.75%. Similarly, in Fig. 14(i), the optimum was obtained with $V_H$ values ranging from 1.4V to 1.8V and $V_L$ =0.6V, which gave the maximum $RSD_{VH}$=0.25%. Thus, the best point for satisfying the abovementioned condition was with $V_H$ =1.4V and $V_L$ =0.6V, allowing maximum $RSD_{VH}$ values of 0.75% and 0.25% for N=8 and N=16, respectively. Equivalently, the maximum $ASD_{VH}$ would be 15mv and 5mv for N=8 and N=16, respectively. In the case study illustrated in Table II and Fig. 9 ($V_H$=2 and $V_L$=1), the maximum $RSD_{VH}$ ($ASD_{VH}$) for satisfying the condition was 0.1% (2mV). Using the optimum values ($V_H$=1.4 and $V_L$=0.6), robustness therefore improves over a factor of 7.5 and 2.5 for $N$=8 and $N$=16, respectively.

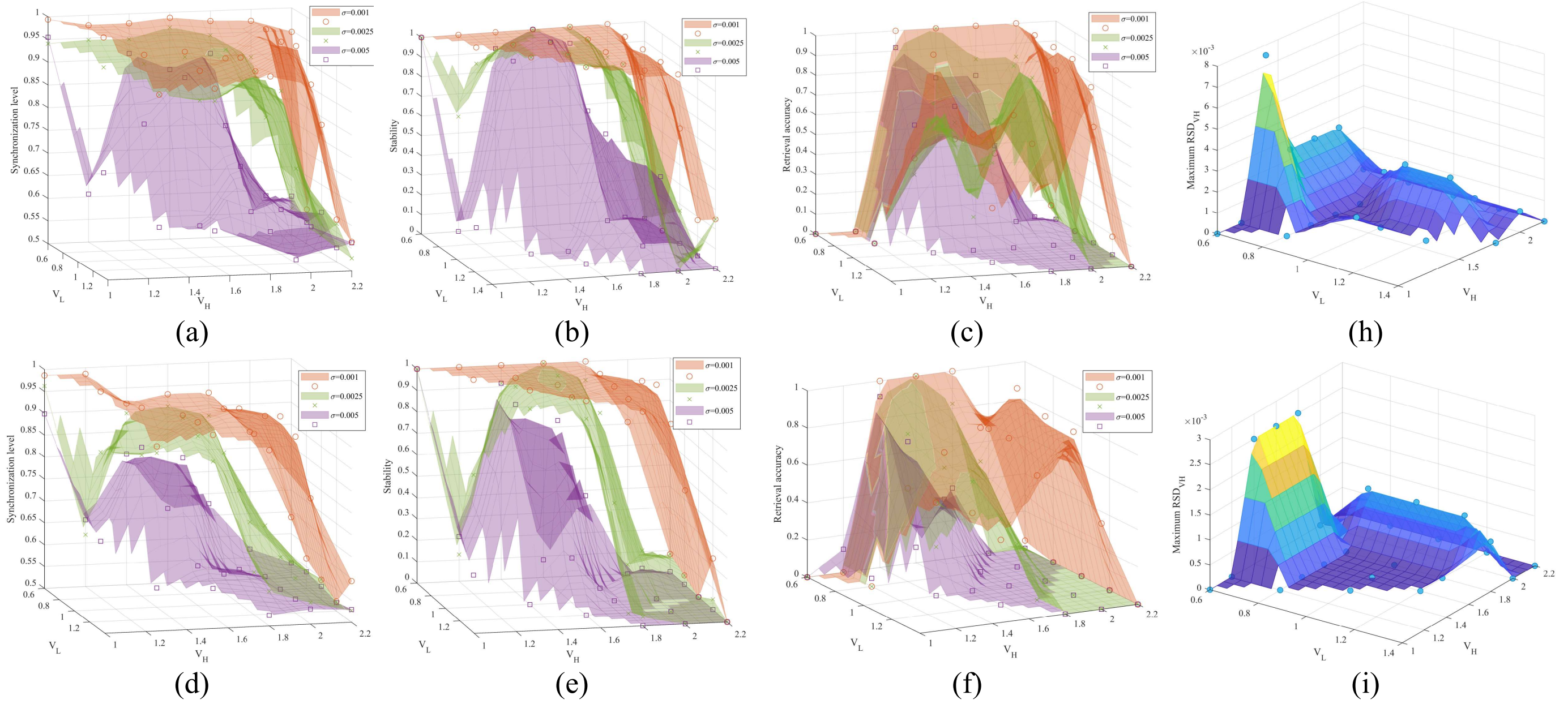

**Fig. 14.** Design space exploration in the presence of mismatches when $V_L$ and $V_H$ are altered. Synchronization levels, stability, and retrieval accuracy for $N$=8 (a-c) and N=16(d-f). Maximum robustness in the design space for $N$=8(h) and $N$=16(i).

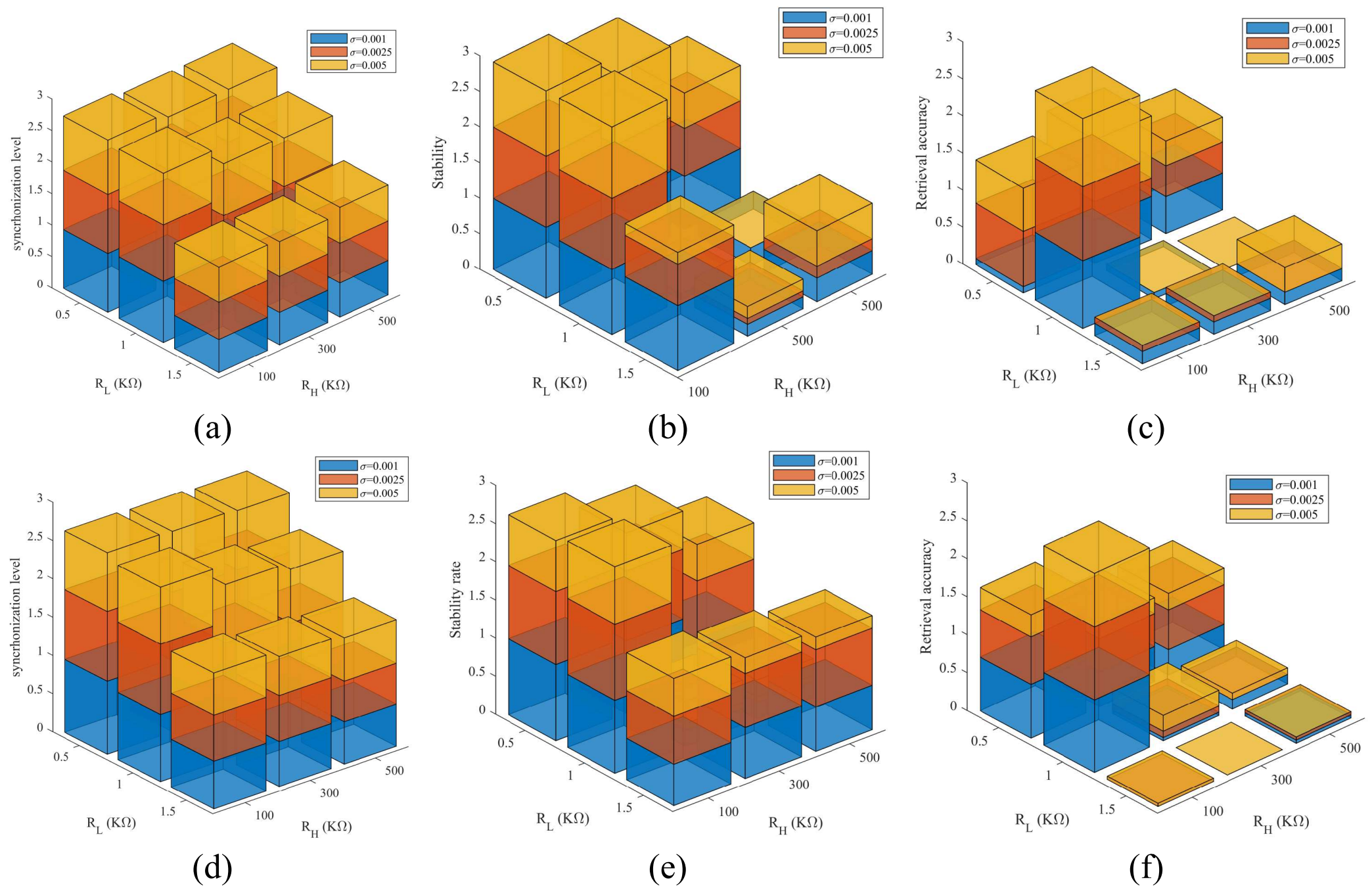

**Fig. 15.** Design space exploration in the presence of mismatches when $R_L$ and $R_H$ are altered. Synchronization level, stability, and retrieval accuracy for $N$=8 (a-c) and N=16(d-f).

In the next step, parameters $R_H$ and $R_L$ were altered and DONN performance was evaluated. Fig. 15 illustrates DONN performance when parameters $R_H$, $R_L$, were varied. $RSD_{VH}$ was changed from 0.001 to 0.005 (three values are shown in the figures). Fig. 15 (a) and Fig. 15 (d) show the synchronization levels for $N$=8 and $N$=16, respectively. The synchronization level did not change much when $R_H$ changed from 100KΩ to 500KΩ. It was reduced, however, when $R_L$ was increased from 0.5KΩ to 1.5 KΩ. Stability (Figs. 12 (b)-(e)) and retrieval accuracy (Figs. 12 (c)-(f) attained maximum robustness at $R_L$=1KΩ and $R_H$=100KΩ.

## V. Conclusion

This paper studies the effect of mismatch on DONN performance. Mismatches were considered in synaptic circuits and differential oscillatory neurons, separately. Memristor variations in the synaptic circuits caused performance degradation in terms of synchronization level, stability, and retrieval accuracy. Simulation results showed that DONNs are tolerant to up to 20% of mismatches in the memristors of the synaptic circuits. Mismatches in the components of differential oscillatory neurons, however, have a more adverse effect on DONN performance. These variations cause non-uniformity in the natural frequency of the differential oscillatory neurons, resulting in desynchronization and instability with 0.5% of natural frequency RSD. It is worth mentioning that as far as the simulation time was reasonable, we increased the number of neurons and altering parameters to improve the generality of the results.

Sensitivity analysis showed that the factor which most affected the non-uniformity of natural frequencies was the high threshold voltage of $VO_2$ devices. It is therefore important to adopt suitable control measures in the fabrication process of $VO_2$ devices, with special attention to high threshold parameter $V_H$, to mitigate the adverse effects of mismatches. In addition, circuit designers can focus on circuit design techniques to reduce the sensitivity of frequency to $VO_2$ devices, specifically parameter $V_H$. Calibration would also be another solution in which calibration techniques (either during device fabrication like laser trimming or adding tunable materials or compact calibration circuit) can be used in this regard. This opens many possibilities and future research avenues, but it is beyond the scope of this paper.

## VI. Appendix

In this section, the $VO_2$ model is reviewed which has been introduced in [18]. Fig. 16 shows the circuit equivalent of the model in which the insulator or metal state is decided by a voltage comparator and is stored in capacitor $C_o$. The output voltage of the comparator ranges from 0 to 1 and is defined by

$$V_o = 0.5(1 + tahn(2\alpha(V^+ - V^-)) \quad (13)$$

where α determines the slope of the comparator transition curve. The device current is given by

$$I = I_f = G_f V_f \quad (14)$$

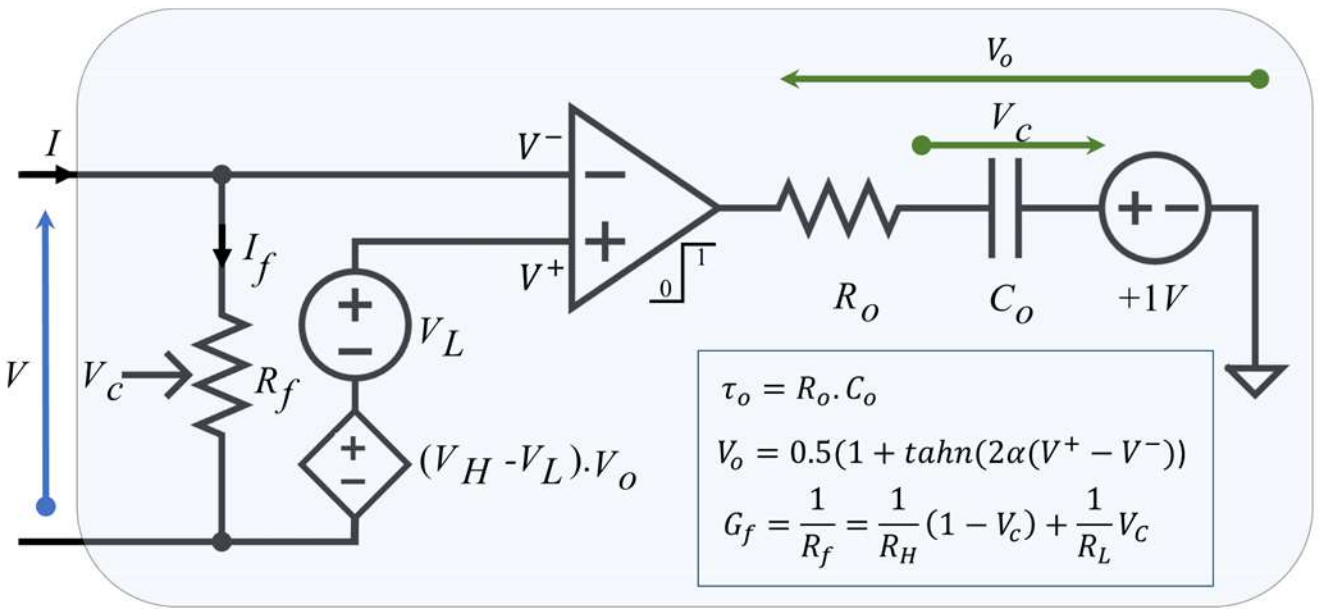

**Fig. 16.** $VO_2$ model [18].

In the insulator state, the comparator output is $V_O$= 1 V, $V_c$= 0, and $R_f = R_H$. By increasing $V$ beyond $V_H$, the comparator output is changed from 1 to 0 which results in the changing of $V_c$ from 0 to 1, and consequently, device resistance $R_f$ changes from $R_H$ to $R_L$ (IMT). In this state (the metal state), if $V$ gradually decreases below $V_L$, the comparator output is changed from 0 to 1 which results in the changing of $V_c$ from 1 to 0, and consequently, device resistance $R_f$ changes from $R_L$ to $R_H$ (MIT). The time constant of the changing of $V_c$ from 0 to 1 (1 to 0) is $\tau_o = R_o.C_o$ which determines the time constant of IMT (MIT).

## ACKNOWLEDGMENT

This work was funded by EU H2020 grant 871501 "NeurONN", and by Spanish Ministry of Economy and Competitivity grant PID2019-105556GB-C31 (NANOMIND) (with support from the European Regional Development Fund).

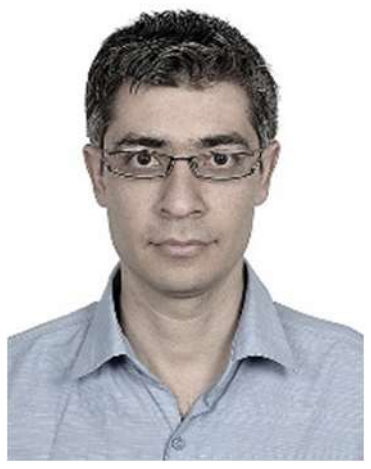

**Jafar Shamsi** received MSc. and Ph.D. degrees in Electrical Engineering from the Iran University of Science and Technology in 2013 and 2018, respectively. His research projects were focused on analog neuromorphic systems including classic and spiking neural networks. In 2020, he joined the Instituto de Microelectrónica de Sevilla (IMSE-CNM) to work on the hardware implementation of oscillatory neural networks as a part of an EU-funded project (NeurONN). He is currently a post-doctoral researcher at the University of Calgary where he is working on the digital implementation of spiking neural networks.

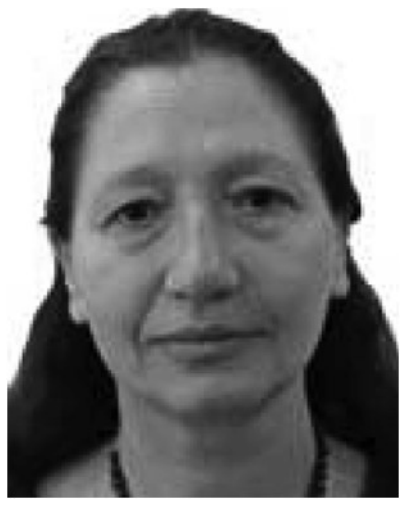

**María J. Avedillo** received her PhD. (summa cum laude) in 1992 from the Department of Electronics and Electromagnetism at the Universidad de Sevilla, which she had joined in 1988 as an Assistant Professor. She has been a Full Professor there since 2010. In 1989, she became a researcher with the Department of Analog Design at the National Microelectronics Center (currently, Instituto de Microelectrónica de Sevilla). She has authored over 150 technical papers in leading international journals and conferences. Her current research interests focus on the design of circuits using emerging devices, including steep slope transistors and phase transition materials, and non-conventional computing paradigms with emphasis on energy constrained applications. In 1994, the Council of the Institution of Electrical Engineers awarded her the Kelvin Premium for two published articles.

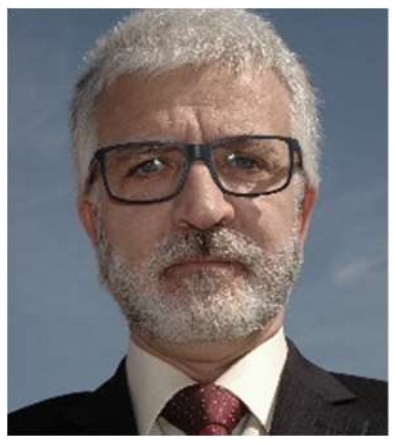

**Bernabé Linares-Barranco** (Fellow, IEEE) received a BSc. degree in Electronic Physics, an MSc. degree in Microelectronics, and a PhD. from the Universidad de Sevilla, Seville, Spain, in 1986, 1987, and 1990, respectively, and a PhD. degree from Texas A&M University, College Station, TX, USA, in 1991. Since 1991, he has been with the Instituto de Microelectrónica de Sevilla (IMSE-CNM), part of the CSIC (Spanish Research Council), where he is currently a Full Research Professor. He has been a Visiting Professor/Fellow with Johns Hopkins University, Baltimore, MD, USA; Texas A&M University, College Station, USA; The University of Manchester, U.K.; and the University of Lincoln, U.K. His recent interests are address-event-representation VLSI, real-time AER vision sensing and processing chips, memristor circuits, and extending AER to the nanoscale. Dr. Linares-Barranco has received two IEEE Transactions best paper awards. From 2011 to 2013, he was the Chair of the Spanish Chapter of the IEEE Circuits and Systems Society. He has been an Associate Editor of IEEE Transactions on Circuits and Systems II: Express Briefs, IEEE Transactions on Neural Networks, and Frontiers in Neuromorphic Engineering.

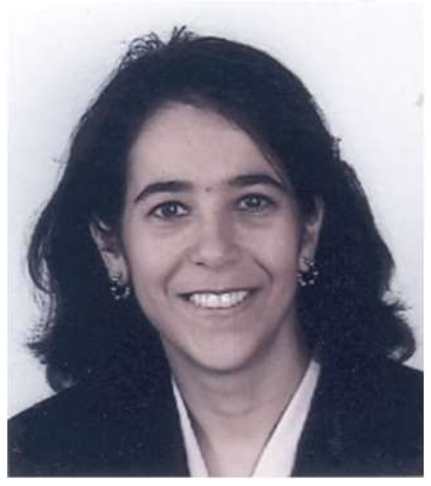

**Teresa Serrano-Gotarredona** (Senior Member, IEEE) received a BSc. degree in Electronic Physics and a PhD. in VLSI Neural Categorizers from the Universidad de Sevilla, Seville, Spain, in 1992 and 1996, respectively, and an MSc. degree from the Department of Electrical and Computer Engineering at Johns Hopkins University in 1997. She was an Assistant Professor at the Electronics and Electromagnetism Department, Universidad de Sevilla, from September 1998 until June 2000. Since September 2000, she has been a tenured scientist at the Instituto de Microelectrónica de Sevilla (IMSE-CNM-CSIC) in Seville, Spain. In July 2008, she was promoted to the position of Tenured Researcher. Since January 2006, she has also been a part-time lecturer at the Universidad de Sevilla. She has published one monographic book, eight book chapters, 69 peer-reviewed journal articles, more than 100 papers in international conferences, and six licensed patents. Her research interests include analog circuit design of linear and nonlinear circuits, VLSI neural-based pattern recognition systems, VLSI implementations of neural computing and sensory systems, transistor parameter mismatch characterization, address-event-representation VLSI, nanoscale memristor-type AER, and real-time vision sensing and processing chips. Dr. Serrano-Gotarredona was co-recipient of the 1995–96 IEEE Transactions on Very Large-Scale Integration (VLSI) Systems Best Paper Award for the paper *A Real-Time Clustering Microchip Neural Engine*. She was also co-recipient of the 2000 IEEE Transactions on Circuits and Systems—I: Regular Papers Darlington Award for the paper *A General Translinear Principle for Subthreshold MOS Transistors*. She is co-author of the book *Adaptive Resonance Theory Microchips*. She has been the Chair of the Sensory Systems Technical Committee of the IEEE Circuits and Systems Society and the Chair of the Spanish Chapter of IEEE Circuits and Systems. She was Academic Editor of the PLoSOne from May 2008 until October 2013, and has also served as Associate Editor for IEEE Transactions on Circuits and Systems—I: Regular Papers and IEEE Transactions on Circuits and Systems—II: Express Briefs. She is currently an Associate Editor of Frontiers in Neuromorphic Engineering.